\documentclass[twocolumn,showpacs,amsmath,amssymb,aps]{revtex4}
\usepackage{graphicx}
\usepackage{epsfig}
\usepackage{longtable}
\begin{document}

\title{$^{49}$Cr: Towards full spectroscopy up to 4 MeV.}
\author{F. Brandolini\( ^{1} \), R.V. Ribas\( ^{2} \), M. Axiotis\( ^{3} \), M. De Poli\( ^{3} 
\), R. Menegazzo\( ^{1} \),  D.R. Napoli\( ^{3} \), P. Pavan\( ^{1} \), J.~Sanchez-Solano\( ^{4} 
\),
 S. Lenzi\( ^{1} \), A. Dewald\( ^{5} \),
 A. Fitzler\( ^{5} \), K. Jessen\( ^{5} \), S. Kasemann\( ^{5} \), P. v. Brentano\( ^{5} \) }
\affiliation{
1 Dipartimento di Fisica dell'Universit\`a  and INFN, Sezione di Padova, Padova, Italy.\\
2 Instituto de Fisica, Universidade de S\~ao Paulo, S\~ao Paulo, Brazil.\\
3 Laboratori Nazionali di Legnaro - INFN, Legnaro, Italy.\\
4 Departamento de F\'{\i}sica Teorica, Universidad Aut\'{o}noma, Cantoblanco, Madrid, Spain.\\
5 Institut f\"ur Kernphysik der Universit\"at zu K\"oln, Germany.}

\date{\today}

\begin{abstract}
The nucleus $^{49}$Cr has been studied analysing $\gamma-\gamma$ coincidences in the reaction
$^{46}$Ti($\alpha$,n)$^{49}$Cr at the bombarding energy
of 12 MeV. The level scheme has been greatly extended at low excitation energy and several new lifetimes have been determined by means of the Doppler Shift Attenuation Method.

  Shell model calculations in the full $pf$ configuration space reproduce well negative-parity  levels. 
Satisfactory agreement is obtained for positive parity levels by extending the configuration space to include a nucleon-hole either in the 1$d_{3/2}$ or in the 2$s_{1/2}$ orbitals.
 A nearly one-to-one correspondence is found between experimental and theoretical levels up to an excitation energy of 4 MeV.
 Experimental data and shell model calculations are interpreted
 in terms of the Nilsson diagram and the particle-rotor model, showing the  strongly coupled nature of the bands in this prolate nucleus. Nine  values of K$^{\pi}$ are proposed for the  levels
observed in this experiment.
 As a by-result it is shown that the  values of the experimental magnetic moments in 1$f_{7/2}$ nuclei are well reproduced without  quenching the  nucleon g-factors.

\end{abstract}


\maketitle

\bigskip


\section{Introduction}

In the last years the nucleus $^{49}$Cr has been studied quite extensively both theoretically
and experimentally.
It was formerly shown that  shell model (SM) calculations 
are able to reproduce the $^{49}$Cr ground state (gs) band and that its rotational
features at low spin can be  described by the particle-rotor model (PRM) as a K=5/2$^-$
band based on the  $\nu$[312]5/2$^-$ Nilsson orbital \cite{Mart2}.
A rotational behavior was already recognized in the most recent  Nuclear
Data Sheets (NDS) compilation \cite{NDS}, where, relying on few experimental levels at low excitation energy, sidebands  with K$^\pi$=1/2$^-$, 3/2$^-$ and 3/2$^+$, based on [321]1/2$^-$, [321]3/2$^-$ and [202]3/2$^+$ Nilsson orbitals, have been suggested.
  More recently, in the frame of a research using heavy ion induced fusion reactions,
   evidence  of two high-K 3-qp rotational bands with
 K\( ^{\pi } \)=13/2\( ^{-} \) and K$^\pi=13/2^+$ has been found \cite{Bra2,Bra3}. They have been interpreted 
 with the prolate Nilsson diagram as due to the excitation of  a proton to the empty [312]5/2$^-$ orbital
 from the [321]3/2$^-$ or the  [202]3/2$^+$ one, respectively, followed by the recoupling to the maximum value of $K$ of the three unpaired nucleons.

SM calculations for natural (negative) parity states were made with the code ANTOINE  in the full {\it pf} configuration space  \cite{Caur} and a very good agreement was obtained.  Good agreement was achieved for the observed  unnatural (positive) parity levels by extending the configuration space to include a hole in the 1d$_{3/2}$ orbital. As B(E2) values are an essential mean for evaluating the nuclear deformation, lifetime measurements with the Doppler shift attenuation method (DSAM) were systematically made.
A recent  review of the SM predictions and of their interpretation for most
  $N\simeq Z$  1$f_{7/2}$ nuclei can be found in Ref. \cite{BraAL}.

It has to be noted, however, that in heavy ion induced fusion reactions  the population of single particle sidebands predicted at low excitation energy is weak. As an example,  in the reaction $^{42}$Ca($\alpha$,n)$^{45}$Ti,
 a  K=1/2$^+$ band, classified as [200]1/2$^+$, was observed  about one MeV above the  [202]3/2$^+$   K=3/2$^+$ band \cite{Ka}. That band was not observed  in a subsequent heavy ion reaction \cite{Be} with a much more efficient set-up.
  The knowledge of non yrast structures is, however, required for a better understanding of nuclear structure. This perspective, often named
 full spectroscopy, became more important recently, also because some properties of  
  non yrast states are fingerprints of nuclear symmetries and supersymmetries \cite{Iac}. Such information was scanty in $\gamma$-spectroscopy, as levels up to about 2.5 MeV had been studied in the  reaction $^{46}$Ti($\alpha,n \gamma)^{49}$Cr  at a bombarding energy of 8 MeV, more than twenty five years ago   \cite{Kas}. We have used the same reaction but, in order to observe non yrast levels in $^{49}$Cr, which are  populated with small cross-section, a high efficiency $\gamma$-detector array was used. The collected experimental data provide a stringent test for modern SM calculations, as non yrast levels are more subjected to residual interactions
 and to the effects of configurations space truncation, due to the increased level density.
 On the other hand,  bands cannot be observed in proximity of their  smooth  terminations, since only states with rather low spin values could be populated. Terminating states are generally well known from heavy-ion induced reactions \cite{BraAL}.

\section{Experimental procedure}

 Excited states were populated in the reaction $^{46}$Ti($\alpha$,n)$^{49}$Cr with the 
  12 MeV \( \alpha  \)-beam  provided by the Cologne FN-TANDEM accelerator.
Five Compton-suppressed Ge detectors and one Compton suppressed CLUSTER
detector were used in the COLOGNE-COINCIDENCE-CUBE-spectrometer \cite{Dewald}. Four
Ge detectors were mounted in the horizontal plane at \( \pm  \) 45$^\circ$,
 \( \pm  \)135$^\circ$. The
fifth Ge detector and the CLUSTER detector were placed along the vertical axis
 below and above the beam line, respectively. 
The target consisted of 1 mg/cm$^{2}$ $^{46}$Ti backed onto 5 mg/cm$^2$ Au. 
A total of 200x10$^6$  $\gamma-\gamma$
 coincidences were collected in 3 days of beam time.

\begin{figure}[b]
\includegraphics[width=0.48\textwidth]{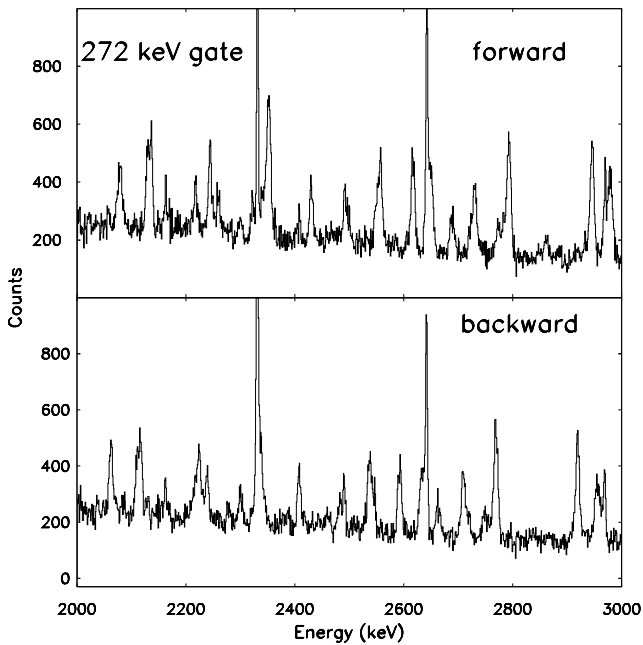}
\caption{ a) Spectra obtained by gating 
on the 272 keV transition in the forward matrix ( upper panel) and in the backward matrix
 ( lower panel).}
\label{fig1}
\end{figure}

\begin{figure*}
\epsfig{file=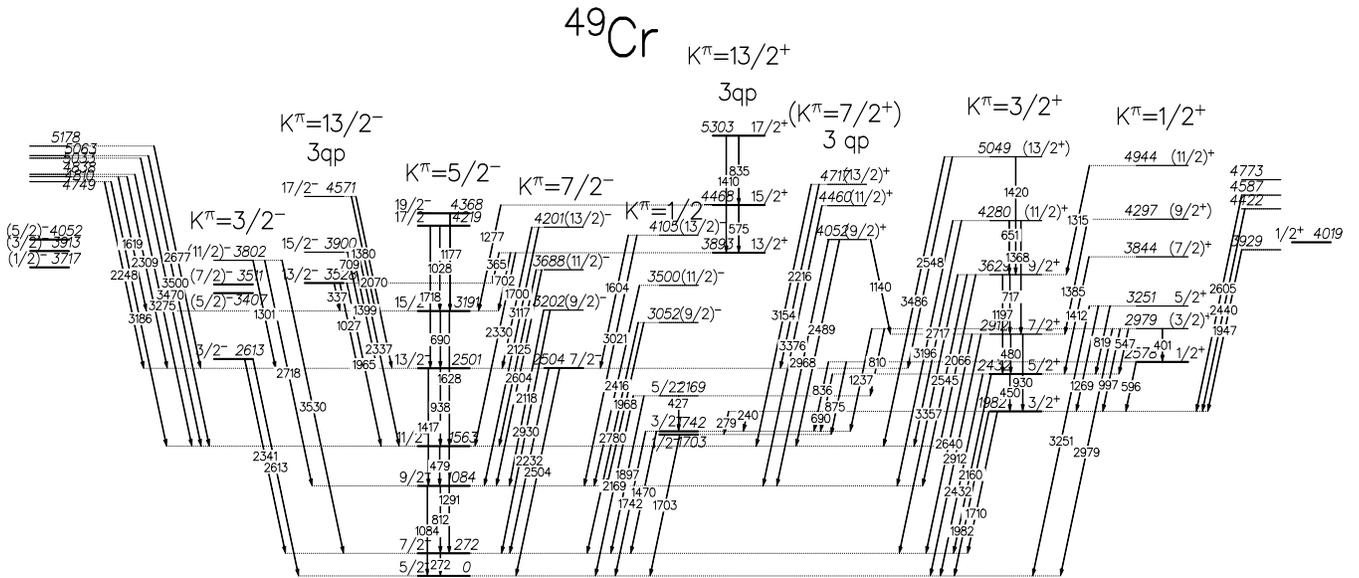,width=18.cm,height=8cm,angle=0}
\caption{ Experimental $^{49}$Cr level scheme.  All known experimental levels are reported up to about 4 MeV. Levels previously known are represented with thicker lines, while, at higher energies, only levels observed in this work are reported. Only observed transitions are reported apart for the ones of
  2613, 2979, 3251 and 2504 keV, taken from Ref. [2].
  The suggested assignments of the  levels 9/2$^-$, 11/2$^-$ and 13/2$^-$ levels to
the bands K=1/2$^-$ and 7/2$^-$ bands may be interchanged.}
\label{fig2}
\end{figure*}

 \begin{figure*}
\includegraphics[width=0.60\textwidth]{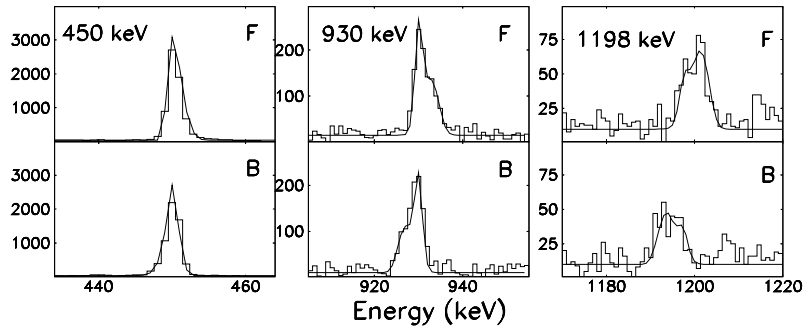}
\protect\caption{ Examples of DSAM lineshape analysis along the K=3/2$^+$ band. }
\label{fig3}
\end{figure*}
 
 \begin{figure*}
\includegraphics[width=0.60\textwidth]{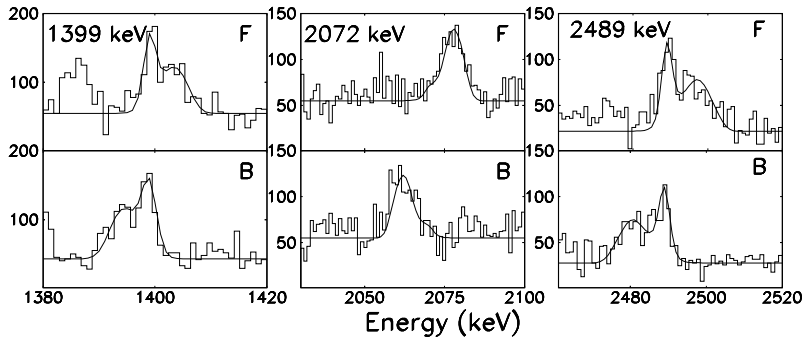}
\protect\caption{ The first two panels show the lineshape analyses of transitions from the 15/2$^-$ and 17/2$^-$ levels of the K$^\pi$=13/2$^-$ band, respectively. The last panel shows the same for a transition from the (9/2$^+$) level attributed to the K$^\pi$=7/2$^+$ band. }
\label{fig4}
\end{figure*}

 About forty percent of them were due to
 the reaction on study. Little less came from  the $^{46}$Ti($\alpha$,p)$^{49}$V 
 reaction, and nearly twenty percent from the inelastic diffusion.
 Since the target sample  contained about five percent of $^{48}$Ti, minor contributions
 came from reactions with this isotope.
  An upper limit was put at about 3.6 MeV in the $\gamma$-ray acquisition.
 This resulted in a limitation for the study of  high lying levels,
  since the production of $\gamma$ rays with an energy
  up to 6.6 MeV is allowed by the kinematics. 
   Owing to the detector geometry no relevant angular correlation information
could be obtained.

 \section{ Data analysis}
 
 \subsection{Level scheme}

  The most useful $\gamma-\gamma$ matrices for extending the level scheme have been the two
 asymmetric ones having all detectors in the first axis and 
 detectors, either at $\pm$ 45 degrees (forward or F matrix) or at $\pm$ 135 degrees (backward or B 
 matrix), in the second axis.
   Many of the observed transitions were either fully Doppler-shifted or not Doppler-shifted, in which cases
  the experimental energy resolution   at about  2 MeV was about 7 keV   and
 3 keV, respectively. The symmetric  $\gamma-\gamma$ matrix was used for the
 analysis of unshifted lines.
 
 A typical analysis of an observed line implied the comparison of spectra
 obtained with the same gate on backward and forward matrices.
  The intrinsic energy and the Doppler shift
 were extracted for all observed transitions by using the most suitable
 gates for each one. The transitions were placed in the level scheme after comparing
 spectra for the same direction, but obtained with different gates, and  successive 
 cross checks. 
   One example for the comparison between forward and backward spectra is shown in
  Fig. 1, where the gate is the 272 keV $7/2^-$$\to$$5/2^-$  transition. The presence of unshifted,
   full shifted and partially shifted lines is evident.
  From the analysis of full shifted lines, one can estimate the average recoil velocity component 
   along the detector axis as 0.45~\% of c, which corresponds to an average
    recoil velocity of 0.65~\% c, in agreement with the kinematics of this reaction.

 The adopted level scheme is shown in Fig. 2. All known levels  are displayed up to 4 MeV.
 The few levels, that were not observed in the present experiment, are not connected by
  transitions.  Above that energy, only levels observed in the present experiment are shown. Levels
for which no new information was obtained for spin and parity are displayed
with thicker lines. Several high spin levels, elsewhere observed, are not reported \cite{Bra2,Bra3}. 
 In particular the gs band had been previously observed up to the band termination at 31/2$^-$ and the K=13/2$^+$ up to   19/2$^+$.

  Transitions to the ground state from levels directly fed by the reaction cannot be
  detected in a $\gamma-\gamma$ coincidence experiment.
 Therefore, some low-spin states could escape an observation if they decay principally  to the ground state.
  This appears to be the case for the K$^\pi$=1/2$^-$, I$^\pi$= 7/2$^-$ level, which is missing  
  in Fig. 2.
 The transitions of the 2613, 2979, 3251 and 2504 keV levels to the gs were taken from Ref. \cite{NDS}.
  Several new levels have been observed above  4 MeV and up to
  5.37 MeV, but some others could have escaped the observation, if the depopulating transitions have 
an energy larger than 3.6 MeV.
 
Observed levels and transitions  are reported in Table I.

\begin{longtable*}[p]{|c|c|c|c|c|l|c|c|c|c|}

\caption{Experimental data for  $^{49}$Cr sidebands. }
\label{table1}\\

\hline
\multicolumn{1}{|c}{ E$_x$} &
\multicolumn{1}{|c} {K-value} &
\multicolumn{1}{|c} {Spin} &
\multicolumn{1}{|c} {$\tau$} &
\multicolumn{1}{|c} {$\tau$} &
\multicolumn{1}{|c} {E$_\gamma $} &
\multicolumn{1}{|c} {BR} &
\multicolumn{1}{|c} {BR} &
\multicolumn{1}{|c} {B(E1)} &
\multicolumn{1}{|c|} {parity} \\

\multicolumn{1}{|c} { } &
\multicolumn{1}{|c} {SM orbital} &
\multicolumn{1}{|c} { }  &
\multicolumn{1}{|c} {present}  &
\multicolumn{1}{|c} {previous$^a$} &
\multicolumn{1}{|c} { } &
\multicolumn{1}{|c} {previous$^a$} &
\multicolumn{1}{|c} {present} &
\multicolumn{1}{|c} {estimate} &
\multicolumn{1}{|c|} {changing} \\ 

\multicolumn{1}{|c|} {(keV)} &
\multicolumn{1}{|c|} { }  &
\multicolumn{1}{|c|} { }  &
\multicolumn{1}{|c|} {(ps)} &
\multicolumn{1}{|c|} {(ps)} &
\multicolumn{1}{|c|} {(keV)} &
\multicolumn{1}{|c|} {\% } &
\multicolumn{1}{|c|} {\% } &
\multicolumn{1}{|c|} {(10$^{-4}$ efm$^2$)} &
\multicolumn{1}{|c|} {mark} \\
\hline
\hline

\endfirsthead

\multicolumn{8}{c}%
{{\bfseries \tablename\ \thetable{} -- continued from previous page}}\\ 
\hline
\multicolumn{1}{|c}{ E$_x$} &
\multicolumn{1}{|c} {K-value} &
\multicolumn{1}{|c} {Spin} &
\multicolumn{1}{|c} {$\tau$} &
\multicolumn{1}{|c} {$\tau$} &
\multicolumn{1}{|c} {E$_\gamma $} &
\multicolumn{1}{|c} {BR} &
\multicolumn{1}{|c} {BR} &
\multicolumn{1}{|c} {B(E1)} &
\multicolumn{1}{|c|} {parity} \\

\multicolumn{1}{|c} { } &
\multicolumn{1}{|c} {SM orbital} &
\multicolumn{1}{|c} { }  &
\multicolumn{1}{|c} {present}  &
\multicolumn{1}{|c} {previous$^a$} &
\multicolumn{1}{|c} { } &
\multicolumn{1}{|c} {previous$^a$} &
\multicolumn{1}{|c} {present} &
\multicolumn{1}{|c} {estimate} &
\multicolumn{1}{|c|} {changing} \\ 

\multicolumn{1}{|c} {(keV)} &
\multicolumn{1}{|c} { }  &
\multicolumn{1}{|c} { }  &
\multicolumn{1}{|c} {(ps)} &
\multicolumn{1}{|c} {(ps)} &
\multicolumn{1}{|c} {keV} &
\multicolumn{1}{|c} {\% } &
\multicolumn{1}{|c} {\% } &
\multicolumn{1}{|c} {(10$^{-4}$ efm$^2$)} &
\multicolumn{1}{|c|} {mark} \\
\hline
\endhead

\hline
\endfoot

\hline \hline
\multicolumn{10}{l}{ a) From Ref. 2, apart for levels at 3528 and 3893 keV [3].}\\
\multicolumn{10}{l} {*) For these transitions a DSAM analysis was performed.}\\
\multicolumn{10}{l}{?) This line could not be observed.}\\
\multicolumn{10}{l}{ $\dag$) Branchings of this level  were evaluated gating on lower transitions.}\\
\multicolumn{10}{l}{ \#) wrongly placed transition.}\\
\endlastfoot


1703&K=1/2$^-$&1/2$^-$ &&$>5$&1703&100&-&&\\
1742&2p$_{3/2}$&3/2$^-$ & $>$1&1.6(5)&1469$^*$&29(3)&-&&\\
&&&&&1742&71(3)&-&&\\
2169&&5/2$^-$&1.5(5)&$>$4&427&-&2.1(6)&&\\
&&&&&1897$^*$&45(3)&44(5)&&\\
&&&&&2169&55(3)&54(5)&&\\
3052&&(9/2)$^-$&$<$0.04&&1968&-&56(5)&$>$6&\\
&&&&&2780&-&44(5)&&\\
&&&&&3052&&?&&\\
3500&&(11/2)$^-$&$<$0.03&&2416&-&100&$>$15&\\
4105&&(13/2)$^-$&$<$0.03&&1604&-&56(5)&$>$28&\\
&&&&&3021&-&44(5)&&\\
2504&K=7/2$^-$&7/2$^-$&$<$0.03&$<$0.012&2232$^*$&33(5)&-&&\\
&1f$_{7/2}$&&&&2504&67(5)&-&&\\
3202&&(9/2)$^-$&$<$0.04&&2118&-&31(6)&$>$7&\\
&&&&&2930&-&69(6)&&\\
&&&&&3202&&?&&\\
3688&&(11/2)$^-$&$<$0.03&&2123&-&31(4)&$>$9&\\
&&&&&2602&&69(4)&&\\
4201&&(13/2)$^-$&$<$0.03&&1700&-&72(4)&$>$30&\\
&&&&&3177&-&28(4)&&\\
2613&K=3/2$^-$&3/2$^-$ & &0.06(2)&2341&59(3)&-&&\\
&1f$_{7/2}$&&&&2613&41(3)&-&&\\
3407&&(5/2)$^-$&&&&-&-&&\\
3511&&(7/2)$^-$&&&3511&52(10)&-&&\\
&&&&&2430&48(10)&-&&\\
3802&&11/2$^-$&0.10(3)&&1301&&15(3)&&\\
&&&&&2718&-&30(4)&&\\
&&&&&3530&-&55(5)&&\\
3528&K=13/2$^-$&13/2$^-$&0.48(8)&0.38(7)&337.2&4.0(5)&4.3(9)&&\\
&1$f^3_{7/2}$&&&&1027.6&16(2)&17(3)&&\\
&&&&&1965.4$^*$&78(3)&79(4)&&\\
&&&&&2444.0&1.6(5)&$<$2&&\\
3900&&15/2$^-$&0.40(7)&&709$^*$&-&35(5)$^\dag$&12&\\
&&&&&1399$^*$&-&35(5)&&\\
&&&&&2337&-&30(5)&&\\
4571&&17/2$^-$&0.20(4)&&352&-&$<$5$^\dag$&6&\\
&&&&&1380$^*$&-&50(7)&&\\
&&&&&2070$^*$&-&50(7)&&\\

1982&K=3/2$^+$&3/2$^+$&$>$2&$>$2.5&240&-&2.4(6)&&E1\\
&$1d_{3/2}^{-1}$&&&&279&18(2)&10(2)&&E1\\
&&&&&1710$^*$&11(2)&6(2)&&M2+E3\\
&&&&&1982&70(2)&81(7)&&E1\\
2432&&5/2$^+$&1.4(4)&1.3$^{+1.2}_{-0.5}$&450$^*$&52(6)&46(6)&25&\\
&&&&&690&-&3.1(6)&0.5&E1\\
&&&&&2160&\#&1.2(4)&0.01&E1\\
&&&&&2432&48(6)&50(6)&0.12&E1\\
2912&&7/2$^+$&0.75(15)&&480&-&33(5)&34&\\
&&&&&930$^*$&-&13(3)&&\\
&&&&&2640$^*$&-&31(5)&0.19&E1\\
&&&&&2912&-&23(4)&0.12&E1\\
3629&&9/2$^+$&0.18(4)&&717$^*$&-&29(5)&27&\\
&&&&&1198$^*$&-&11(2)&&\\
&&&&&2066&-&4.0(6)&0.167&E1\\
&&&&&2545&-&25(4)&0.53&E1\\
&&&&&3357&-&30(5)&0.28&E1\\
4280&&11/2$^+$&0.30(6)&&651$^*$&-&28(4)$^\dag$&21&\\
&&&&&1368$^*$&-&42(5)&&\\
&&&&&2717&-&17(3)&0.18&E1\\
&&&&&3196&-&13(2)&0.08&E1\\

5049&&(13/2$^+$)&$<$0.1&&769&-&$<$10$^\dag$&&\\
&&&&&1419&-&21(3)&&\\
&&&&&2548$^*$&-&38(4)&&(E1)\\
&&&&&3486$^*$&-&41(5)&&(E1)\\

2578&K=1/2$^+$&1/2$^+$&$>$1&&596&-&12(3)&&\\
&$1d_{3/2}^{-1}$&&&&836$^*$&62(5)&57(7)&$<$6&E1\\
&&&&&875$^*$&38(5)&30(5)&$<$3&E1\\
2979&&(3/2$^+$)&$>$1&&401$^*$&-&59(8)&&\\
&&&&&547$^*$&-&100&&\\
&&&&&810&-&40(5)&$<$3.9&E1\\
&&&&&997$^*$&-&27(4)&$<$1.4&E1\\
&&&&&1237&-&25(4)&$<$0.7&E1\\
&&&&&2979&-&?&&E1\\
3251&&5/2$^+$&0.20(5)&&819$^*$&-&44(6)&17&\\
&&&&&1269$^*$&77(5)&44(6)&6&\\
&&&&&2979&\#&$<$5&$<$0.04&E1\\
&&&&&3251&23(5)&12(3)&0.07&E1\\
3844&&(7/2$^+$)&0.30(6)&&1412$^*$&&100&&\\
&&&&&3572&&?&&(E1)\\
&&&&&3844&&?&&(E1)\\

4297&&(9/2)$^+$&0.05(2)&&1385$^*$&&100&&\\
&&&&&4025&&?&&	(E1)\\
4944&&(11/2)$^+$&0.07(2)&&1314$^*$&&100&&\\
&&&&&3860&&?&&E1\\

3893&K=13/2$^+$&13/2$^+$&$>$10&&364.4&22(2)&17(3)&$<$2&E1\\
&$1d^{-1}_{3/2} 1f^2_{7/2}$&&&&701.9&18(2)&18(3)&$<$0.3&E1\\
&&&&&2330.0&60(2)&65(4)&$<$0.03&E1\\
4052&K=(7/2$^+$)&(9/2$^+$)&0.26(4)&&1140&&10(2)&2&\\
&$1d^{-1}_{3/2} 1f^2_{7/2}$&&&&2489$^*$&&27(4)&0.5& (E1)\\
&&&&&2968$^*$&&63(6)&0.7& (E1)\\
4460&&(11/2+)&0.23(4)&&3376&&100&0.8& (E1)\\
4717&&(13/2$^+$)&0.70(10)&&2216$^*$&&56(8)&0.5& (E1)\\
&&&&&3154$^*$&&44(8)&0.1 &E1\\
&others&&&&&&&&\\
4749&&&$<$0.05&&2248&&41(8)&&\\
&&&&&3187$^*$&&59(8)&&\\
4810&&&$<$0.05&&1619$^*$&&44(8)&&\\
&&&&&2309&&56(8)&&\\

\end{longtable*}


Data related to the gs K=5/2$^-$ and the K=13/2$^+$ bands are not reported since they were already discussed in detail in Ref. \cite{Bra2,Bra3}. 
In the same Table the branching ratios are also displayed, to which errors a systematic contribution of 10 \%  has been added to the statistical one, in order to account for angular correlation effects. When possible, they were obtained by gating on a transition directly feeding the level of interest.
It is explicitly indicated when gates on transitions below of the branches were used, which lead to larger uncertainties. 
 A question mark was put for the branches which could not be measured, in which 
 case the sum of all branching ratios could not be normalized to 100 percent.
Some variations with respect to  NDS are worth  noting: a) The decay from levels 15/2$^-$ 
at 3900 keV and 13/2$^+$ at 3893 keV were  mixed up in Ref. 
\cite{Cam}. b) The  2160 keV branch of the 2432 keV level is very small, in agreement with Ref. \cite{Kas}
  and in disagreement with NDS \cite{NDS}. c) The previously reported branch of 2979 keV from the 3251 keV level   \cite{NDS} is not observed. Most probably that line is produced by the  transition of the 2979 keV level to gs. 

Experimental mixing ratios are not discussed here, since the few reported ones  have a large 
uncertainty \cite{NDS}. 
 Due to the experimental conditions, only a tentative spin-parity assignment
 was made for some levels. The proposed spin-parity and K assignments will be
justified later.

\subsection{Lifetimes and electromagnetic reduced rates}

For DSAM lifetime determinations, the program LINESHAPE  has been used \cite{Well} and the
Northcliffe-Schilling stopping power \cite{NS}, corrected for atomic shell
effects \cite{Sie}, was adopted. 
Spectra gated from transitions below the ones examined were used ( i.e. the standard procedure) since
the experimental conditions did not allow the use of the NGTB procedure, which does
not depend on the sidefeeding time of the examined level \cite{BRibas}.
In this work  the sidefeeding was assumed to occur instantaneously at
the reaction time. This is corroborated by the observed large number of full shifted lines.
 Examples of DSAM analysis are shown in Fig. 3-4.

 Obtained lifetimes for bands not yet studied \cite{Bra1,Bra2} are  reported in Table I, while the deduced B(E2) and B(M1) values are shown in Table II. In order to check the reliability of the presently obtained values, the lifetime of the 13/2$^-$ yrast level was re-evaluated to be 0.17(2) ps, in agreement with the previously obtained value of 0.15(2) ps \cite{Bra3}.

 In order to determine the level parities the upper limit (UL) of
 3$\cdot10^{-4}$ W.u., extracted from  data for several nuclei in this region in Ref. \cite{Bra2,Bra3,Bra1,Bra4,Bra7,Bra8,Bra9},  was adopted for E1 transitions.  Such criterion was found to be very successful in these works. The B(E1) value of most relevant transitions are reported in the last but one column of Table I in units of $10^{-4}$ e$^2$fm$^4$, which is approximately one W.u..
 In the case that all transitions from a levels lead to the gs band, only the largest estimate or limit is reported. If the B(E1) value exceeds the UL the parity change is excluded. If B(E1) value is lower, the M1+E2 character cannot be excluded, but in some cases is unfavored so that E1 multipolarity may be tentatively proposed. E1 assignments are reported in the last column of Table 1.

 \section{Discussion}

\subsection{ Particle rotor model.}

In a previous work a deformation parameter $\beta\simeq$0.26 was deduced for the low part of the  $^{49}$Cr gs-band  \cite{Bra2}.
 The most used formulae for describing the properties of deformed nuclei are those related to the rigid axial rotor.

 For the electromagnetic (em) moments, they are well known in the case of a definite value of the spin projection K (rotor model)  \cite{BM}.
 
  The intrinsic electric quadrupole moment Q$_\circ$ is related to the spectroscopic one Q$_s$ by the
relation:

\begin{equation}
 Q_s=Q_\circ{3K^2-I(I+1) \over (I+1)(2I+3)}
 \label{eq=1}
 \end{equation}

 The intrinsic quadrupole moment can also be derived from the B(E2) values, where it
  is usually denoted as  Q$_t$:
 
 \begin{equation}
 B(E2)={5 \over 16\pi}Q_t^2<I_iK20|I_fK>^2
 \label{eq=2}
 \end{equation}
   
Concerning the magnetic properties, the g-factor of a level with  K$\neq$1/2
is related to the collective g-factor g$_R$ and to the intrinsic value g$_K$ by the formula:

\begin{equation}
 g=g_R + (g_K-g_R) {K^2 \over I(I+1)}
 \label{eq=5}
 \end{equation}
 For M1 transitions one has similarly:
 
 \begin{equation}
 B(M1)= {3 \over 4 \pi} <I_iK10|I_fK>^2 (g_K-g_R)^2 K^2 \mu_N^2
\label{eq=6}
 \end{equation}

 The B(E2) values within a band are sensitive to the K-value 
 via the Clebsh-Gordan coefficient. This is also the case of  Q$_s$,
 which, however, is rarely known experimentally. 
 The value of the deformation parameter is assumed to be given by the formula 
  $Q_t= 1.09 Z A^{2/3}\beta(1+0.36\beta) \,\, {\rm f m}^2$
  which accounts for nuclear volume conservation in the case of deformation \cite{Lo}.

  Eq. (3) and (4) show that the magnetic properties are sensitive to the nature
   of the involved quasiparticles.

 These formulae provide mostly a qualitative interpretation tool, since the assumption of a pure value of K is not valid in general. 
 In fact, even under the extreme hypothesis that the unpaired neutron does not
 interact singularly with the other nucleons,  one has to account for the coupling of
 the spin of the neutron with the rotational moment in order to conserve the total angular
 moment I. This is made by the
 particle-axial rotor model, in brief particle-rotor model (PRM),
  in which context K is not anymore a good
 quantum number because of the Coriolis force.

\begin{figure*}
\includegraphics[width=0.90\textwidth]{49crfig5.ps}
\protect\caption{ Particle-rotor predictions in $^{49}$Cr assuming axial
symmetry and $\beta$=0.26. }
\label{fig5}
\end{figure*}

 Since the standard PRM does not account for residual
 interaction of the valence nucleon with the other nucleons, it is less 
  reliable than SM, but it may provide a structural interpretation of the SM calculations.
  
    It should moreover be  remarked that, approaching 4 MeV of excitation,
     the increase of the level density and the decrease of the 
  deformation, may give rise to large configuration mixing.


Fig. 5 shows the level scheme  predicted by the PRM at low excitation. 
The bands are identified with the dominant Nilsson configuration. For PRM calculations the code  of Ref. \cite{Rag} was used, which is able also to describe
the coupling of a particle with a triaxial rotor.
 The standard set of parameters for an axial symmetry were here adopted
together with  the 2$^+$ level energy of
the $^{48}$Cr core, without any further adjustment of parameters.

\begin{longtable*}[p] {@{\extracolsep{\fill  }}|l|l|l|l|l|l|l|l|l|l|l|l|}

\caption{Experimental and SM reduced rates. Parity non changing. }
\label{table2}\\

 \hline
 {\ \ \ \ \ Line \ \ \ \ }& \ E$_\gamma$\  &\ E$_\gamma$\ &\ $\gamma$-BR\ &\ $\gamma$-BR\ &
\ \ $\tau$\ &
 $\ \tau$\  &\ B(E2)\ &\ B(E2)\ &\ B(E2)\ &\ B(M1)\ &\ B(M1)\ \\
 
  & exp. & SM & adopted & SM & adopted & SM & rotor & exp. &
  SM & exp. & SM\\

  & keV & keV & \% & \% & ps & ps & $e^2fm^4$ & $e^2fm^4$ &
  $e^2fm^4$ & $\mu^2_N$ & $\mu^2_N$\\
\hline
\hline
\endfirsthead

\multicolumn{7}{c}%
{{\bfseries \tablename\ \thetable{} -- continued from previous page}}\\ 
 \hline
  \ \ \ \ \ Line & E$_\gamma$ & E$_\gamma$ & $\gamma$-BR & $\gamma$-BR & $\tau$ &
  $\tau$ & B(E2) & B(E2) & B(E2) & B(M1) & B(M1)\\
 
  & exp.  & SM & adopted & SM & adopted & SM & rotor & exp. &
  SM & exp. &  SM \\

  & keV & keV &  \% &  \%  &  ps  &  ps  &  $e^2fm^4$  &  $e^2fm^4$  &
  $e^2fm^4$ & $\mu^2_N$ & $\mu^2_N$\\
\hline
\endhead

\hline
\endfoot

\hline \hline
\multicolumn{11}{l}{The numbers in subscript to spin values refer to the calculated levels in Fig. 6.}\\
\multicolumn{12}{l}{ In the case that the sum of branching ratios is less than 1, the theoretical lifetime}\\
\multicolumn{12}{l}{ has to be compared with the experimental one divided by the sum of branching ratios. }\\
\multicolumn{11}{l} { $^*$ Data for the observed 7/2$^-$ level, assigned to the K=7/2 $^-$ band,  are inserted }\\
\multicolumn{11}{l}{in the K=1/2$^-$ band  only for a comparison.   }\\
\multicolumn{11}{l}{  $^\dagger$ The sum of branchings may be less than 1 if there is a E1 branch to 
the gs.   }\\
\endlastfoot

$K=1/2^-$   &  &   &   &   &   &   &   &   &   &   &\\
$1/2\to5/2$&1703&1464&100 &100&$>$5&27&&$>$0.4   &2.1&&\\
$3/2\to5/2$&1742&1472&71(3)&75  &1.6(5) &12&&&3.25 & &0\\
$3/2\to7/2$&1470&1192&28(3)&25& & & & &2.28& &-\\
$3/2\to1/2$&39&8&$<$0.1&0&&&234&&241&&0.165\\
$5/2_2\to5/2$&2169&1873&54(3)&55&1.5(5)&2.3&&6.1&4.12&0.002&0\\
$5/2_2\to7/2$&1897&1594&44(3)&45&&&&9.7&2.5&0.002&0.001\\
$5/2_2\to1/2$&465&409&$<$0.1&&&&233&&252&&-\\
$5/2_2\to3/2$&427&401&2.1(5)&&&&65&&69&0.009&0.001\\
$7/2_2\to5/2$&(2504)$^*$&2299&&&$<$0.012&0.21&&&0.3&&0.011\\
$7/2_2\to7/2$&&2020&&&&&&&1.5&&0.012\\
$7/2_2\to9/2$&&1114&&&&&&&4.9&&0.013\\
$7/2_2\to3/2$&&826&&&&&240&&306&&-\\
$7/2_2\to5/2_2$&&425&&&&&25&&32.9&&0.160\\
$9/2_2\to5/2$&3052&2753 &0&4&$<$0.04&0.37&&&0.3&&-\\
$9/2_2\to7/2$&2780&2474 &44(6)&20&&&&&3.6&&0\\
$9/2_2\to9/2$&1968& 1569&56(6)&68&&&&&51&&0\\
$9/2_2\to5/2_2$&(883)&880 &0&8&&&300&&350.2&&-\\
$9/2_2\to7/2_2$&548&455 &&0&&&46&&17.9&&0\\
$9/2_2\to7/2_3$&548&228&&0&&&&&0.2&&0\\
$K=7/2^-$&&&&&&&&&&&\\
$7/2_3\to5/2$&2504&2525&67&79&$<$0.012&0.003&&&52.1&&0.981\\
$7/2_3\to7/2$&2232&2246&33&21&&&&&17.9&&0.372\\
$7/2_3\to9/2$&1420&1340&$<$5&0&&&&&0.5&&0.004\\
$7/2_3\to5/2_2$&335&652&$<$5&0&&&&&0.14&&0.001\\

$9/2_3\to5/2$&3202&3147&0&0&$<$0.03&&&&3.4&&-\\
$9/2_3\to7/2$&2930&2868&69(6)&73&&&&&0.5&&0.529\\
$9/2_3\to9/2$&2118&1963 &31(6)&26&&&&&5.1&&0.500\\
$9/2_3\to5/2_2$&1033&1275&0&0&&&253&&1.9&&-\\
$9/2_3\to7/2_2$&(698)&850 &0&0&&&&&1.4&&0.008\\
$9/2_3\to7/2_3$&(698)& 622&0&0&&&279&&185.5&&0.534\\

$K=3/2^-$&&&&&&&&&&&\\
$3/2_2\to5/2$&2613&2435&59(3)&68&0.06(2)&0.18&&&2.7&&0.012\\
$3/2_2\to7/2$&2341&2156&41(3)&32&&&&&20.3&&-\\

$5/2_3\to3/2_2$&(894)&842&-&&&&340&&5.2&&0.142\\

$7/2_4\to5/2$&3511&3380&52(10)&45&&0.008&&&4.6&&0.071\\
$7/2_4\to7/2$&3239&3100&$0$&15&&&&&16.8&&0.018\\
$7/2_4\to9/2$&2430&2195&48(10)&40&&&&&20.0&&0.186\\
$7/2_4\to3/2_2$&(898)&0&0&&&&179&&67.8&&-\\
$7/2_4\to5/2_3$&(104)&0&0&&&&269&&38.1&&0.096\\

$K=13/2^-$&&&&&&&&&&&\\
$13/2_2\to11/2$&1965&1815&78(3)&64&0.42(6)&&&&1.1&0.010&0.004\\
$13/2_2\to13/2$&1027&846&16(2)&32&&&&&0.8&0.025&0.014\\
$13/2_2\to15/2$&337&225&4.0(5)&4&&&&&14.3&0.12&0.044\\

$15/2_2\to11/2$&2337&2313&35(5)&25&0.40(7)&&&11.6&11.2&&-\\
$15/2_2\to13/2$&1399&1345&35(5)&50&&&&&20.5&0.016&0.040\\
$15/2_2\to15/2$&709&753&30(5)&25&&&&&40.3&0.118&0.146\\
$15/2_2\to13/2_2$&372&499&$<$5&&&&&&137.6&&0.012\\

$17/2_2\to13/2$&2070&2098&50(7)&40&0.20(4)&0.34&&50&24.7&&-\\
$17/2_2\to15/2$&1380&1506&50(7)&60&&&&&2.0&0.054&0.038\\
$17/2_2\to13/2_2$&(1045)&1251&0&0&&&&&27.4&&-\\
$17/2_2\to15/2_2$&(673)  &752&0&0&&&&&152&&0\\

$K=3/2^+$&&&&&&&&&&&\\
$5/2\to3/2$&450&547&46(6)&&1.4(4)&3.3&340&&284.4&0.178&0.070\\

$7/2\to3/2$&930&905&13(3)&38&0.75(15)&2.6&179&203&175.7&&-\\
$7/2\to5/2$&480&358&33(5)&62&&&269&&247.6&0.220&0.125\\

$9/2\to5/2$&1198&1098&11(2)&43&0.18(4)&0.68&269&202&210.2&&-\\
$9/2\to7/2$&717&740&29(5)&57&&&176&&88.0&0.148&0.127\\

$11/2\to7/2$&1369&1148&42(5)&79&0.30(6)&0.53&319&330&256.4&&-\\
$11/2\to9/2$&651&408&28(4)&&&&123&&107.0&0.159&0.081\\

$K=1/2^+$&&&&&&&&&&&\\
$1/2\to3/2$&596&544&12(3)&&$>$1&55&&&124.8&&0.0016\\

$3/2_2\to3/2$&997&1068&15(2)$\dagger$&73&$>$1&0.76&&&17.6&&0.055\\
$3/2_2\to5/2$&547&521&54(6)$\dagger$&11&&&&&54.6&&0.052\\
$3/2_2\to1/2$&401&524&31(5)$\dagger$&16&&&251&&164.8&&0.179\\

$5/2_2\to3/2$&1269&1314&44(6)&85&0.30(6)&0.12&&&21.9&0.041&0.101\\
$5/2_2\to5/2$&819&840&44(6)&15&&&&&0.8&0.152&0.132\\
$5/2_2\to1/2$&(673)&770&0&0&&&251&&194.6&&-\\
$5/2_2\to3/2_2$&(272)&246&0&0&&&72&&88.9&&0.393\\

\end{longtable*}

  Apart of the fact that the PRM model cannot predict the observed 3-qp bands, there is correspondence for most observed low lying levels up to
about 4 MeV.

  The assumption of collective triaxiality could not reproduce the size of the observed gs-band signature
splitting. Cranked shell model (CSM) calculations predict some signature splitting for a collective triaxiality, i.e.  with a negative sign for the deformation parameter $\gamma$ in the Lund convention, but its size is also insufficient. 

  A further limitation of the particle-rotor model is that it cannot
 explain the backbending of the gs band  at I= 19/2, which can be interpreted as a  termination
  of  $\nu=3$ configurations and thus as  an effect of the competition of the seniority scheme
 with rotational collectivity \cite{BraSev,Juo2}.

 Other negative parity bands built with {\it pf} configurations  are expected above 4 MeV,
as for example those based on the [301]1/2$^-$, [301]3/2$^-$ and [303]5/2$^-$ orbitals.  Due to the configuration mixing, 
   interband transitions  are often predicted to prevail over intraband ones, owing to the larger transition energies, qualitatively  explaining why few intraband transitions have  been observed.

  In Table III  Q$_s$ and g values calculated with PRM are compared with the
  rotor predictions of  Eq. (1) and (3).
    The latter values were  obtained with the  PRM   by multiplying the Coriolis coupling term by a null factor. 
   In this way the decoupling factor is added to Eq. (3)   for K$^\pi$=1/2$^-$, giving rise to staggering.
     The predicted level scheme keeps, anyhow, the strongly coupled appearance.
    In Eq. (3) and (4) g$_R$ =0.5 is taken and the suggested 0.6 
    quenching factor is assumed for the nucleon g-factor \cite{Rag}. 
    A particularly strong mixing  occurs between levels  close in energy, as 
    in the case of the head of the K=7/2$^-$ band, near to the 7/2$^-$ level with K=1/2, in which case the
   PRM g-factor value is slightly positive while the rotor one is negative.
   
 Concerning positive parity levels, ``extruder'' bands K=3/2$^+$ and 1/2$^+$ are expected at low excitation energy, being based on the [202]3/2$^+$ and [200]1/2$^+$ orbitals, respectively. Such bands  have been observed at low energy in the nucleus $^{45}$Ti, where they have been satisfactorily described by   PRM \cite{Ka}. 
   The 3/2$^+$ level at 1982 keV is strongly excited by $\ell_n$=2 pick-up in $^{50}$Cr.  This agrees with the PRM prediction that the [202]3/2$^+$ orbital has a nearly pure spherical 1d$_{3/2}$  hole-configuration. Similarly, the yrast 1/2$^+$ level at 2578 keV is clearly the band-head of the K=1/2$^+$ band based on the [200]1/2$^+$ hole configuration as it is strongly excited  by $\ell_n$=0 pickup reaction on $^{50}$Cr \cite{NDS} but, in  this case, the orbitals [200]1/2$^+$ and [211]1/2$^+$ share the  $\ell_n$=0 component  to a comparable amount. 
 
 The first positive parity shell model orbital above the Fermi level is the 1g$_{9/2}$ one. In the deformed nucleus $^{49}$Cr the lowest intruder level of positive parity is expected to be the
  9/2$^+$ one belonging to the decoupled band based on the $\nu$[440]1/2$^+$  orbital. Since this level can be qualitatively described with the configuration $^{48}$Cr$\otimes\nu g_{9/2}$ it should be  strongly excited in a neutron stripping reaction, but
  this is not feasible since $^{48}$Cr is unstable.
  In $^{51}$Cr a level is observed  at an  excitation energy of 4.16 MeV in the $^{50}$Cr(d,p) reaction with a large $\ell_n=4$ spectroscopic factor of $S_n=3.2$ and it is thus described with a dominant  $^{50}$Cr$\otimes\nu g_{9/2}$ configuration.
   The expected  band is, however, not yet observed with $\gamma$-spectroscopy.
 Since the deformation of $^{50}$Cr is comparable with  that   of $^{48}$Cr \cite{Bra1}, the decoupled band  in $^{49}$Cr  should start above 5 MeV, because the sloping up  $\nu$[312]5/2$^-$ orbital is empty in $^{48}$Cr. We estimate thus that  it is out of the sensitivity range of  the present experiment. 
  

 \subsubsection{Questions related to isospin}
 
  It is  worth noting that the hole excitations,  giving rise to the bands K=3/2$^+$ and 1/2$^+$, are not of pure neutron as assumed in  PRM, since isospin conservation implies a  proton-hole  contribution of one third for the yrast states 3/2$^+$ and 1/2$^+$. 
  
   Since the lowest T=3/2, I=7/2$^-$ state, isobaric analogue state (IAS)  of the $^{49}$V gs, lies at 4764 keV, in the following the comparison of  $^{49}$Cr levels with SM theoretical levels will be limited to states of the lowest isospin T=1/2.

Few comments are anyhow added on higher IAS's.    The T=3/2,  I=3/2$^+$ IAS of the yrast one in $^{49}$V, lies at 5573 keV in $^{49}$Cr  \cite{Fu}. In this reference, the sum of the  $\ell_n$=2 pick-up strengths of the 3/2$^+$ IAS's in $^{45}$Ti and $^{49}$Cr was probably somewhat overestimated because  some contribution of the [202]5/2$^+$ orbital, expected at similar energies, was likely included.
  Concerning the 1/2$^+$ levels, a large experimental $\ell_n$=0 pick-up strength is concentrated on a level at 6470 keV, which  was interpreted as the T=3/2 IAS  of the yrast 1/2$^+$ in $^{49}$V.
 The   $^{49}$Cr 1/2$^+$ level with configuration [211]1/2$^+$  is predicted more than 3 MeV above the one based on  the [200]1/2$^+$ orbital, but it is likely fragmented owing to the high level density, while IAS's are  more robust against mixing.
It may be that some $\ell_n$=0 strength due to [211]1/2$^+$ orbital was attributed
 to IAS fragmentation of the T=3/2,  I=1/2$^+$ state, leading also in this case to an overestimate of the sum of spectroscopic factors with isotopic spin T$_>$ \cite{Kas}.

\subsection{Shell Model calculations.}

\subsubsection{ Negative parity}

Negative parity levels have been calculated with the code ANTOINE \cite{Caur}, using the
KB3G residual interaction in the full {\it pf} configuration space \cite{Poves}. Five states for each spin value were calculated from 1/2 to 21/2, making sure that all levels up to 4 MeV are included.
All experimental energy levels up to 4 MeV are displayed in Fig. 6 for a comparison with  SM
predictions. It appears that all of them can be related with a theoretical 
 level and that, in the other way around, only few predicted levels cannot be 
 related with an experimental one.   The tentative spin-parity assignments, reported in brackets in Fig. 6, rely in part on the predictive capability of SM calculation. 
    One has to note, however,  that  the theoretical levels of the K=1/2$^-$ band lie about 270 keV below
  the experimental ones.
  
\begin{figure*}[htbp]
\includegraphics[width=0.95\textwidth]{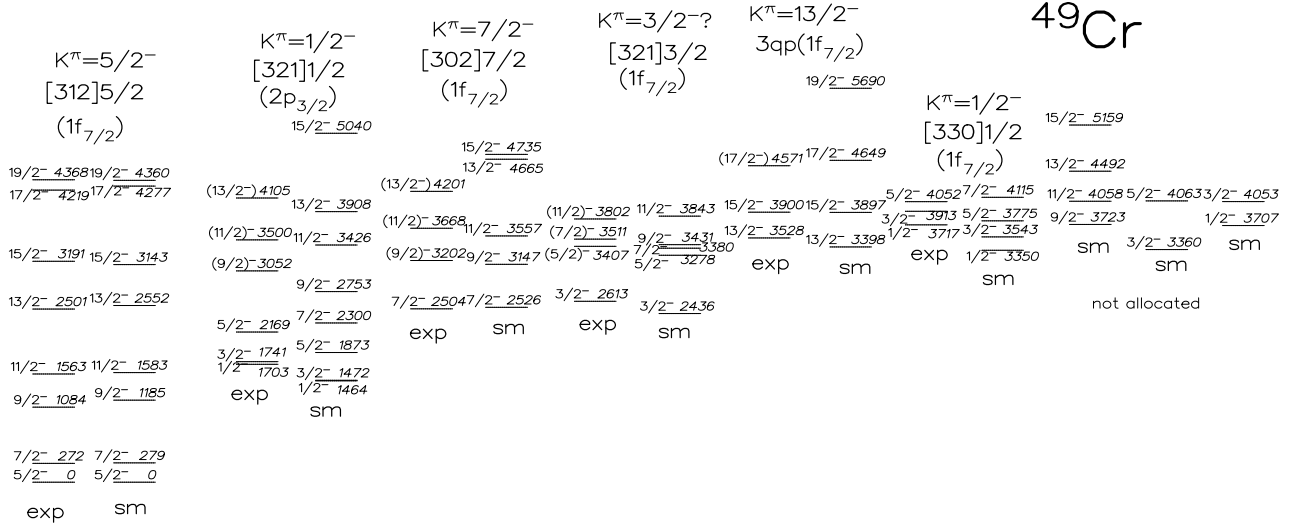}
\protect\caption{Comparison of experimental negative parity levels with SM
predictions. }
\label{fig6}
\end{figure*}

 SM  B(E2) and B(M1) values are compared to the experimental
ones in Table II. In the same Table also the B(E2) rotor values of Eq. (2) are
reported.
The branching to the gs of the (9/2)$^-$ levels assigned to the bands with K$^\pi$=1/2$^-$ and 7/2$^-$ are assumed to be negligible on the basis of the SM estimates.

\begin{table}
\caption{Theoretical  Qs and g-factor values in single-particle bands. 
 $g_s^{eff}=0.6 g_s$ was assumed in PRM and ROT evaluations.}

\label{table3}

\begin{tabular}{|c| c| c| c| c| c| c| c|} 

\hline
  K   &Spin &Q$_s^{prm}$ &Q$_s^{rot}$ & Q$_s^{sm}$  &g$^{prm}$ & g$^{rot}$&
  g$^{sm}$ \\
  \hline
  \hline
  5/2$^-$ &5/2 & 36.1  &35&36.3    &-0.27&-0.14&-0.20 \\
       &7/2 &11.6 &8.2&9.6   &0.14&-0.07&-0.01  \\
       &9/2 &-6.8 &-8.7&-8.4   & 0.02&0.27&0.23 \\
  1/2$^-$ &1/2 &    0 &0&  0 &1.11&1.13 & 0.848    \\
       &3/2 &-21.5&-20&-22.3 &-0.24&0.02&-0.23 \\
       &5/2 &-30.8&-29&-32.8  &0.55&0.73&0.47  \\
       &7/2 &-17.4&-33&-36.6  & 0.04&0.28&0.20 \\
       &9/2 &-33.6&-36&-40.6    & 0.42&0.64&0.43  \\
   7/2$^-$&7/2 & 23.1&46&31.7   &0.04&-0.15&0.10  \\
       &9/2 &3.7  &18&8.3   & 0.23 &0.08&0.28 \\
       &11/2&-9.7&1&-1.0&0.28&0.21&0.41\\
   3/2$^-$&3/2 & 22.0&19&22.5 &-0.08 &-0.17&-0.42 \\
       &5/2 & -4.6  &-8.1 &-14.6&0.36  &0.23 &0.83\\
       &7/2 & -16.8  &-22.5 &-9.3& 0.48 &0.43 &0.34\\

  3/2$^+$ & 3/2 &21.5&22.5&21.4&0.552&0.57&1.06\\
       & 5/2 &-8.7 &8.0&12.2&0.501&0.52&0.74\\
       & 7/2 &-24.0&-22.5&-14.3&0.497 &0.51&0.749\\
  1/2$^+$ & 1/2 &0 &0&0&-0.554&-0.56&-1.06\\
       & 3/2 &-21.5 &-22.5&-20.2&0.682&0.67&0.66\\
       & 5/2 &-31.4 &-32.1&-20.0&0.312&0.29&0.54\\
\hline
\end{tabular}

\end{table}

 The decay towards the gs band of the levels 1/2$^-$, 3/2$^-$ and 5/2$^-$ of the K=1/2$^-$ band
  is  correctly predicted to be very small, in accordance with the K-selection rule.
      
The 7/2$^-$ level at 2504 keV is not assigned
 to the  K=1/2$^-$ band, but to the K$^\pi$= 7/2$^-$ band based on the [303]7/2$^-$ orbital, because of its very fast decay.   Its  lifetime is, in fact, quoted to be shorter than 12 fs \cite{NDS}, while
  an upper limit of 30 fs is found in the present  experiment. 
 As reported in Table II, this agrees with the theoretical prediction
of 3 fs for the lifetime of the  K$^\pi$=7/2$^-$ band head, while the alternative
assumption (K$^\pi$=1/2$^-$) would lead to the prediction of a too long lifetime of about 200 fs.
The  B(M1) values for transitions of the K$^\pi$=7/2$^-$ band head to the gs band are predicted to be of the order of one $\mu_N^2$, in contrast with the very small ones of the low spin members of the K=1/2$^-$ band, indicating that  no K-selection rule is active. One can conclude that the wavefunction of the observed 7/2$^-$ state has a dominant K=7/2 component. 

A further spectroscopic tool is provided by the SM predictions for Q$_s$ and g-factor values reported in  Table III.
 They have in fact to be considered reliable estimates, on the basis of
the  agreement achieved for the level scheme and for the B(E2)  and B(M1) values.
In this context, the suggestion that the observed yrast 7/2$^-$ level is a band-head
  is confirmed by the large positive Q$_s$ value of 31.7 efm$^2$ calculated by SM (see eq. 1). 
The SM g-factor value is  small and positive, as in the PRM calculations. Since the rotor value is negative 
 some mixing is confirmed. No quenching of nucleon g-factors is assumed in the SM calculations,
 as it  will be  justified later.

 While the lowest terms of the bands are firmly established and well reproduced by SM calculations, some uncertainty
 remains for higher terms, quoted in brackets, whose predicted em properties are rather poor.  
 While SM em moments agree  with rotor properties and thus with the K-hindered decay from the K=1/2 band, the observed decays of levels 9/2$^-$ and 11/2$^-$ of the two bands do not exhibit  peculiar experimental differences, in disagreement with SM predictions.
  This is explained by the fact that, if one increases  by 270 keV the theoretical values for the K=1/2$^-$ band in order to compensate the  energy offset with respect to the experimental values, the theoretical energies of the levels 7/2$^-$, 9/2$^-$, 11/2$^-$ and 13/2$^-$ of the bands K=1/2 and 7/2 get close, so they  likely mix strongly. In this way the K=1/2 members may acquire from the K=7/2- band a sizable M1 strength to the gs band. The adopted band assignments  merely correspond to a slightly better correspondence between theoretical and experimental levels. A possible ambiguity is not explicit in Fig. 6. One of the two (11/2)$^-$ levels could be the 9/2$^-$ level corresponding to that predicted at 3431 keV.

The 3/2$^-$ level at 2613 keV is  predicted by SM at 2436 keV, but the   Q$_s$ value for the suggested head of the
K=3/2$^-$ band  has the opposite sign of  the rotor estimate. The SM g-factor value has a negative sign which reveals its neutron character, but the size is -0.40, i.e. about 3 times bigger than the rotor predictions  for a 1$f_{7/2}$ neutron.
 It is moreover predicted to be connected to a calculated 1/2$^-$ level at 3707 keV with B(E2)= 83 efm$^2$.  This band does not have clear rotational features.  This can be related to  K-mixing, which would be consistent with the  observed strong signature splitting of the levels (5/2)$^-$ at 3407 keV  and (7/2)$^-$ at 3511 keV 
  (these levels  were observed in transfer reaction with $l_n$=3, so that the experimental assignment for both is 5/2$^-$-7/2$^-$). The 11/2- level  observed at 3802 keV corresponds probably  to the calculated one
 at 3843 keV, which is predicted to be strongly connected to the  (7/2)$^-$ level  at 3511 keV.
  

The yrare 13/2$^-$ level was already discussed in Ref. \cite{Bra2},
 where  it was identified  as the head of a
  K=13/2$^-$ band. SM predicts in fact Q$_s$ =55 efm$^2$. According to the rotor prediction of eq. 1, 
its deformation would be about 20\%  lower than that of the gs band. Its calculated g factor is 0.77, which is about twice that  of the corresponding yrast level, and indicates the 2-proton alignment.
 A semiclassical estimate is obtained considering the sum of the projection of the unpaired nucleons magnetic moment along the symmetry axis: $ g_{13/2}= 2/13[ 1.4(5/2+3/2)-0.40\cdot 5/2]=0.71$, which agrees with the  SM values.
  A reduction of rotational collectivity is suggested by the lower contribution of the 2p$_{3/2}$ orbital with respect to the gs band, calculated by SM. In fact it has been shown  that the 2p$_{3/2}$ occupation is the essential ingredient that leads to deformation \cite{Caur}.
The breaking of a proton pair also reflects into a theoretical g-factor value larger that the rotor
one.   The levels (15/2$^-$) and (17/2$^-$) belonging to the K=13/2$^-$ band are well
  characterized by their decay scheme.

A  K=9/2$^-$ band is predicted to be based on a 9/2$^-$ level at 3723 keV. Its Q$_s$  value is  55.7  efm$^2$ and it is connected to the 11/2$^-$ levels at 4058 MeV with a B(E2) value  of 279 e$^2$fm$^4$. 
 The band continues with levels 13/2$^-$ and 15/2$^-$ but its nature is not yet understood.

\subsubsection {Positive parity}
Experimental and theoretical data for positive parity levels are reported in Table I and II, together with those of negative parity levels.
  The experimental  levels are compared with SM predictions in Fig. 7,  where the excitation energy of the yrast 3/2+ is adjusted to the experimental value and where also the 21/2$^+$ and 23/2$^+$ reported in Ref. \cite{Bra2} are shown.
The experimental bands  with K=3/2$^+$ and K=13/2$^+$ have been already discussed in that reference but now  the K=3/2$^+$ band has been  substantially extended. They have  been  reproduced there with SM calculations, where one nucleon was lifted from the 1$d_{3/2}$ orbital and  three
particles were allowed to be promoted from the 1$f_{7/2}$ orbital to the rest of the $pf$ configuration space. 
 In this frame, the band K=13/2$^+$ is described with a $\pi d_{3/2}^{-1}\otimes ^{50}$Mn(K=5,T=0) configuration, where a parallel coupling occurs. This band should terminate at 33/2$^+$, while levels are  seen only up to 23/2$^+$.

\begin{figure}[htbp]
\epsfig{file=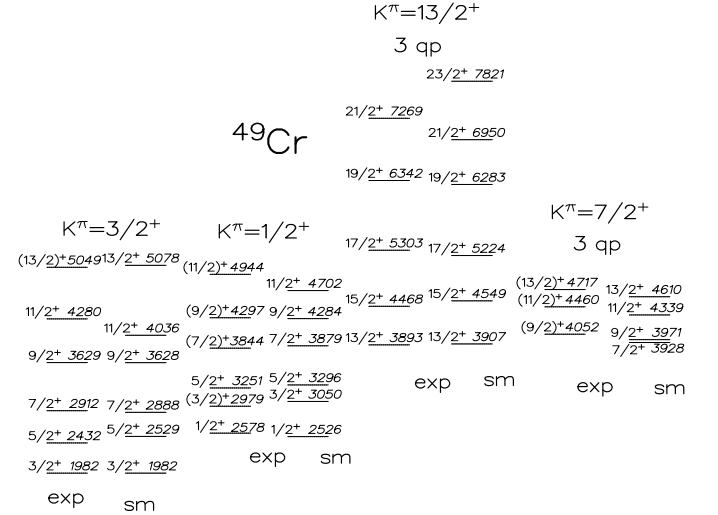,width=9.cm,angle=0}
\protect\caption{Comparison of positive parity levels with SM predictions. }
\label{fig7}
\end{figure}

 The same calculations  predict, however,  the yrast 1/2$^+$ two MeV too high.
  This  is caused by the configuration space truncation, which does not account for
the large contribution of the 2s$_{1/2}$ spherical orbital in the Nilsson orbital [200]1/2$^+$. This does not affect much the description of the
[202]3/2$^+$ orbital, which is calculated by particle-rotor model to have a nearly pure 1d$_{3/2}$ configuration.


\begin{figure}[htbp]
\epsfig{file=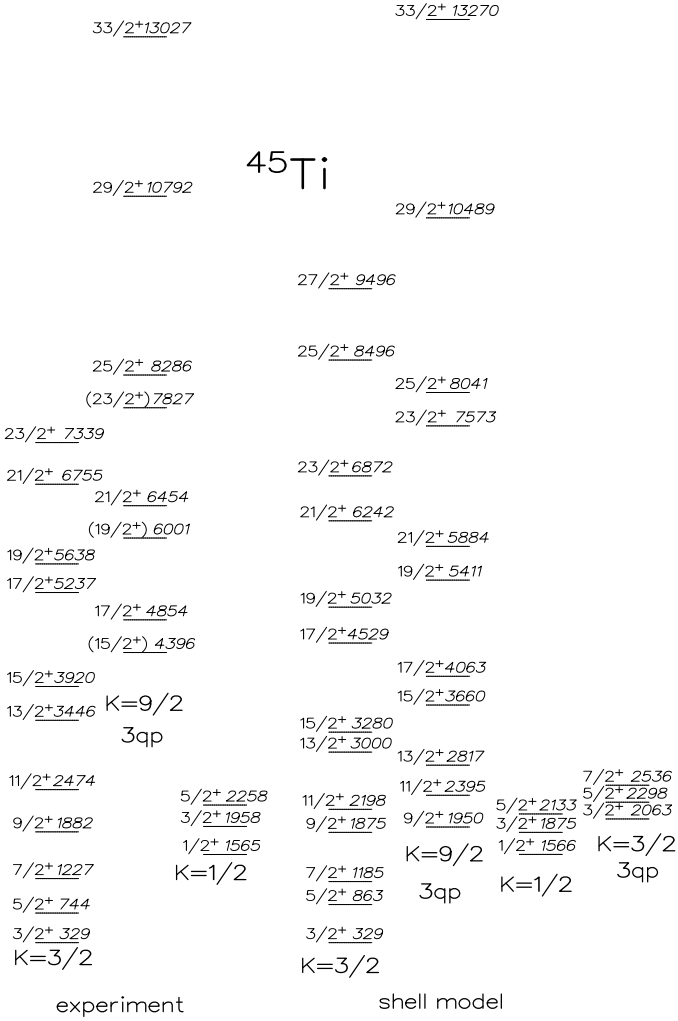,width=9.cm,angle=0}
\protect\caption{Comparison of positive parity levels with SM predictions in $^{45}$Ti. }
\label{fig8}
\end{figure}

It has been thus necessary to allow a hole also in the 2s$_{1/2}$ orbital.
The interaction was similar to that used  for the K=3/2$^+$ band in $^{47}$V  \cite{PovesV}, with nearly  standard values of binding energies. Satisfactory agreement is generally obtained for the level energies, while the B(M1) values of some K=1/2 band members are predicted too large. 
  The members of the K$^\pi$=1/2$^+$ band are not compared with theory in Table II above 5/2$^+$ because  branchings to low members of the gs band could not be observed.
 A 3-qp band with K=7/2$^+$  is  predicted by SM  in the case of an  antiparallel coupling of spins.
  Its decay to the K=3/2$^+$ one is predicted to occur mainly via E1 transitions to the gs band.
  There are few candidates to belong to such a band:
 i.e. the levels at 4052, 4460 and 4717 keV that have  rather long lifetimes in spite of the high energy of the depopulating $\gamma$-rays.
 The level at 4052 keV is candidate to be  9/2$^+$ by its peculiar decay.
If this is the case, the band-head 7/2$^+$ is very close in energy, but its decay to the
first excited level was not observed.

A confirmation of the existence of core-excited 3-qp bands at low energies is provided by the recent observation in $^{45}$Ti of a  sequence from 17/2$^+$ up to  33/2$^+$ \cite{Bed}, which can be  described as the upper part of the K=9/2$^+$ band due to the parallel alignment of three particles. Such a band can be described with the configuration $\pi d_{3/2}^{-1}\otimes ^{46}$V(K=3,T=0), which terminates at 33/2$^+$ as the one in $^{49}$Cr. The reason why termination was not seen in $^{49}$Cr  may be that a hole excitation require nearly 2 MeV more. In $^{51}$Mn an intruder band 1g$_{9/2}$ was observed rather than core-excited bands \cite{Ek}.
The comparison of  SM  predictions with observed positive parity levels in $^{45}$Ti is  shown in Fig. 8. Differently to Ref. \cite{Bed}, a level  (15/2$^+$) is located at 4396 keV, by inverting the cascade following the decay of the 17/2$^+$ level. A 3 qp band with K=3/2 is predicted but not yet observed.

\subsection{ About  effective nucleon g factors.}

 Bare values of the nucleon g-factors are adopted in this paper, while 
in an early work \cite{Bra1} and in recent papers, effective values of the nucleon g-factor were
used. In particular in Ref. 
\cite{Neu} and \cite{Poves} the following effective parameters have been adopted in the shell $pf$:
 $g_s^{eff}=0.75g_s^{bare}$, $g_{\ell} (\pi)^{eff}$=1.1 and $g_{\ell}(\nu)^{eff}$= -0.1, which were justified   mainly based on electron induced   M1-excitation data \cite{Neu}.
One must, however, consider that g-factor values provide a  better test for the assessment of effective
values. In fact, in some cases the experimental values are very precise and  SM calculations predict very well the level scheme, while in the case of M1 
excitation one may suspect uncertainties in evaluating the experimental cross section and the sum rule of the B(M1) values, which range over a large energy interval where the quality of the SM calculations is not always good.
 
 \begin{table}[h]
\caption{Comparison of experimental and theoretical g-factors. }
\label{table4}

\begin{tabular}{|c| c| c| c| c|} 

\hline
nucleus&level& g$_{exp}$& g$_{SM}^{bare}$& g$_{SM}^{eff}$\\
\hline
\hline
$^{49}$V&7/2$^-$&1.277(5)&1.248&1.179\\
$^{51}$V&7/2$^-$&1.4710(5)&1.442&1.350\\
$^{51}$Mn&5/2$^-$&1.4273(5)&1.360&1.280\\
$^{53}$Mn&7/2$^-$&1.435(2)&1.386&1.332\\
&&&&\\
\hline
\end{tabular}

\end{table}

In Table IV the comparison is limited to the odd-Z nuclei $^{49}$V, $^{51}$V, $^{51}$Mn and $^{53}$Mn, for which the SM predictions are very good at low excitation energy.
 Assuming the bare nucleon g-factors in the calculations,  one gets an average precision of 5\%, while using   the effective values  the agreement gets considerably poorer. It is inferred that in this  major shell, where  account  is taken for both the dominant $\ell$ +1/2  and the conjugated $\ell$-1/2 orbitals, there is  no experimental evidence of the meson exchange currents (MEC) effects on the nucleon g-factors, discussed in Ref. \cite{Cast}.
   In this context, the quenching of nucleon g-factor, which is usually applied to reduce
 the disagreement with experimental values of PRM calculations,
  appears to account roughly for the configuration mixing among the shell model orbitals.
  
  A similar comparison for odd N nuclei  would be not conclusive since the effect of a quenching is small.
  The present conclusions confirm that of a recent paper \cite{Hon}, where the value $g_s^{eff}$=0.9 $g_s^{bare}$ was derived from a comparison with SM of 113 experimental magnetic moments in the mass range A=47-72.
 In the {\it sd} major shell the evidence of the need of using effective values for the nucleon
  g-factors to reproduce the experimental g-factors is also not firm \cite{Wild} and similar effective values as in Ref.  \cite{Hon} could be adopted. It is not yet understood, however,  why the sum rule in M1 excitation processes  requires, as mentioned, the use of strongly quenched $g_s$ values.

One  reason why quenching of nucleon g-factor operators due to MEC was often assumed is  related
to the   quenching of the $G_A$
coefficient in Gamow Teller (GT) weak decay, which was recently deduced to be also of about 0.75 \cite{Poves}. The weak and the em decays occur, in fact, via similar operators, since the GT operator is $\vec \sigma\tau$, while the magnetic moment operator is
 $\vec \mu = g_\ell ^{is}\vec l +1/2 g_s^{is} \vec \sigma +g_\ell ^{iv} \vec l \tau _z +1/2 g_s ^{iv} \vec \sigma
\tau_z$, where the upper scripts $is$ and $iv$ refer to the isoscalar and isovector terms, respectively.
 The smallness of the global MEC effects on $g$-factor values  was ascribed  in Ref. \cite{Hon} to the compensation of the quenching of the  operator $\sigma\tau$ with the  MEC effects  on the other operators, but it was observed that the operator $\vec \sigma\tau$ may experience different MEC effects in em and weak interactions \cite{Cast,Fu2}.


\section {Conclusions.}

The level scheme  adopted for $^{49}$Cr includes several new levels, which are displayed as
thinner lines in Fig. 2.
  K$^\pi$ values  for nine sets of levels observed in this
 experiment are proposed.  Six bands are described by the  Nilsson configurations lowest in energy, where the K-values 
  suggested in Nuclear Data Sheets \cite{NDS} are confirmed for three of them.
  A clear correspondence is established between particle-rotor  and
shell model calculations. 
 Moreover three 3-qp bands are observed: the K=(7/2$^+$) is new, the K=13/2$^+$ is confirmed and the K=13/2$^-$ is  substantially extended.  
 SM  calculations in the full $pf$ configuration space
account  for all observed negative-parity levels up to about 4 MeV. 
Calculations account reasonably  well for all observed positive  parity 
levels, extending the configuration space to include a nucleon-hole either in the 1d$_{3/2}$
 or in the 2s$_{1/2}$ orbitals. 

 A comprehensive description of $^{49}$Cr is presented. Since, however, 
 some assignments are tentative, a measurement 
 with a larger $\gamma$-detector array would be desirable in order to verify the proposed scenario.
 A measurement in coincidence with neutrons would, moreover, allow to determine precisely
 also the branches to the ground state, probably revealing the missing
 7/2$^-$ level of the K=1/2$^-$ band. 
Particularly challenging is  to improve the knowledge of the 3-qp K=(7/2$^+$) band. 

 The principal achievement of this work is to show  that full spectroscopy is at hand and that it can be fruitful not to leave the still fertile stability valley for cultivating friable slopes.
 As a by-result, it is shown that in shell model calculations the bare values of nucleon
g-factor are suitable for calculating magnetic effects.



\end{document}